\documentstyle[12pt]{article}

\newfont{\klein}{cmbx9 scaled 1200}
\newfont{\name}{cmbx10 scaled 1095}
\newfont{\namex}{cmbx12 scaled 1200}

\begin{document}

\setcounter{page}{0}

\begin{minipage}[t]{14cm}

\begin{center}

\today
 
 Constraint Quantization of Slave-Particle Theories
 
Christian Helm, Joachim Keller \\
Institut f\"ur Theoretische Physik 
der Universit\"at Regensburg,  93040 Regensburg, Germany\\
 
\end{center}


We start from the Barnes-Coleman slave-particle description, where the 
Hubbard operators $X$ are decomposed into a product of fermionic ($f_{\alpha}$)
and bosonic ($b$) operators. The quantum mechanical constraint 
$b^{\dagger} b + \sum_{\alpha} f_{\alpha}^{\dagger} f_{\alpha} = 1$ 
is treated within the framework of Dirac's method for the quantization of 
classical constrained systems. This leads to modified algebraic properties 
of the fundamental operators: $b b^{\dagger} b = b$, $f_{\alpha} 
f_{\beta}^{\dagger} f_{\gamma} = \delta_{\alpha \beta} f_{\gamma}$ and 
$ f_{\alpha} b^{\dagger}= 0  $. Thereby the algebra of the $X$-operators is 
preserved exactly on the operator level. Matrix representations 
of the above algebra are constructed and a resolvent-like perturbation 
theory for the single-impurity Anderson model is developed.

\begin{tabular}{rl}
keywords: slave-boson,
          constraint quantization,
          resolvent perturbation theory
\end{tabular}
 
\vspace*{1cm}
 
Christian Helm \\
Institut f\"ur Theoretische Physik der 
 Universit\"at Regensburg,
93040 Regensburg,
Germany, FAX: +49-941-943-4382 \\
e-mail: christian.helm@physik.uni-regensburg.de

\end{minipage}

\newpage




In systems like the single-impurity Anderson-model
\begin{eqnarray}
H = \sum_{k \sigma} c^{\dagger}_{k \sigma} c_{k \sigma} +
    \epsilon_f \sum_{\sigma=1}^N X_{\sigma \sigma} + \nonumber  \\
    \sum_{k \sigma=1}^N V_k ( X_{0 \sigma} c^{\dagger}_{k \sigma} + 
			  X_{\sigma 0} c_{k \sigma} )
\end{eqnarray}
the doubly occupied state of the strongly correlated f-electron 
($N$: spin degeneracy) is forbidden due to a strong local Coulomb repulsion. 
This is described by the so-called Hubbard operators
 $X_{\alpha \beta} = \mid \alpha \rangle \langle \beta \mid$
($\alpha, \beta = 0, \dots N$), which 
fulfill a projector-like algebra and a completeness relation
\begin{equation}\label{xop}
X_{\alpha \beta} X_{\gamma \delta} = \delta_{\beta \gamma} X_{\alpha \delta}
\; \; \; \; 
 \sum_{\alpha=0}^N  X_{\alpha \alpha} = {\bf 1}_{ {\cal H}_{\rm phys} } .
\end{equation}
In the standard slave-boson-approach \cite{slave} the $X$-operators are 
decomposed into a product of fermionic ($f_{\alpha}$) and bosonic ($b$) 
degrees of freedom
\begin{equation}
X_{\alpha \beta} = \psi_{\alpha}^{\dagger} \psi_{\beta}
\; \; \; \;
\psi_{\alpha}^{\dagger} = ( b^{\dagger}, f^{\dagger}_{\sigma} ) , 
\end{equation} 
which are intuitively interpreted as creators and annihilators of empty ($b$)
and occupied ($f_{\sigma}$) states. 
In order to eliminate unphysical degrees of freedom, an 
additional constraint has to be implemented in the 
functional integral:
\begin{equation}\label{constraint}
( b^{\dagger} b + \sum_{\sigma} f_{\sigma}^{\dagger} f_{\sigma} - 1  )
 \mid {\rm phys} \rangle = 0 
\end{equation}
In the usual treatment of slave-boson-theories  this constraint is 
only fulfilled on a mean-field level and it is difficult to control 
the contribution of unphysical parts of the Hilbert space. 

Motivated by this  we aim at a decomposition 
$X_{\alpha \beta} ( t ) = \psi_{\alpha}^{\dagger}( t ) \psi_{\beta}
( t )$ of $X$-operators into a product 
of two operators with own dynamics while fulfilling the algebra and 
completeness relation of equ. \ref{xop} exactly on the operator level. 

The constraint equ. \ref{constraint} can be treated exactly by Dirac's 
method \cite{dirac} for the quantization of classical constrained systems, 
where the standard (anti)-commutation relations of the constrained degrees 
of freedom are modified in order to incorporate the constraint on the 
operator level. This procedure has also been
 carried out (with inconsistent 
final results) in \cite{slavedirac}. The corrected  result reads as 
\begin{equation} \label{slave}
\psi_{\alpha} \psi_{\beta}^{\dagger} \psi_{\gamma} = \delta_{\alpha \beta}
  \psi_{\gamma} , \, \, \,  
 \sum_{\alpha} \psi_{\alpha}^{\dagger} \psi_{\alpha} ={\bf 1}_{\cal H} .
\end{equation}                         
Note that these algebraic properties are equal for all $\psi_{\alpha}$,
which are neither bosons nor fermions, but that  
 equ. \ref{xop} is reproduced exactly.  

One  possibility to fulfill this algebra is to enlarge 
the $N+1$-dimensional physical Hilbert space ${\cal H}_{\rm phys}$ by an 
unphysical "vacuum" state $\mid {\rm vac} \rangle$ of the slave-particle 
operators: $\mid 0 \rangle = b^{\dagger} \mid {\rm vac} \rangle$ and 
$\mid \sigma \rangle = f_{\sigma}^{\dagger} \mid {\rm vac} \rangle$.  
Then on the space ${\cal H} = {\cal H}_{\rm phys} \oplus 
{\tilde {\cal H}}$ 
\begin{equation} 
\psi_{\alpha}^{\dagger} := (0, \dots, e_{\alpha}^{\dagger} )  
\in {\bf C}^{(N+2) \times (N+2)} 
\end{equation}
with $(e_{\alpha})_{n} := \delta_{n, \alpha}$ $n = 0, \dots, N$. 
Thereby the usual resolvent theory is recovered, as the dynamics on the 
unphysical space ${\tilde {\cal H}}$ is reduced to the identity: 
$\psi_{\alpha} (t) = {\bf 1}_{\tilde {\cal H}}\psi_{\alpha} e^{- i H t}$.  
 
In addition to this  we constructed  all possible matrix
 representations of the 
algebra of equ. \ref{slave}, which can serve as a  starting point 
of alternative perturbational approaches beyond the usual resolvent 
expansion. 

In this paper we present for simplicity only 
 representations on Hilbert spaces ${\cal H} := {\tilde {\cal H}}
\otimes {\cal H}_{\rm phys}$, which are a tensor product 
of an arbitrary separable Hilbert space
 ${\tilde {\cal  H}}$ and the physical Hilbert space 
${\cal H}_{\rm phys}$ and which  turn out to show interesting new features.
 In this case  $\psi_{\alpha}^{\dagger}$ can be represented as
\begin{equation}
\psi_{\alpha}^{\dagger} = ( 0, \dots, 0, a_{1 \alpha}^{\dagger}, \dots, 
a_{k \alpha}^{\dagger} ) 
\in {\bf C}^{k (N+1) \times k (N+1)} 
\end{equation}
with $a_{i \alpha} a_{j \beta}^{\dagger} = \delta_{ij} \delta_{\alpha \beta}$,
$a_{i \alpha} \in {\bf C}^{k (N+1)}$ for $i,j=1, \dots,k$ and $\alpha,\beta 
= 0, \dots, N$. Without loss of generality we choose the special orthonormal 
 basis ${ ( a_{i \alpha} ) }_p := \delta_{p, i (N+1) + \alpha +1 }$.
Consequently, $\psi_{\alpha} \psi_{\beta}^{\dagger} = \delta_{\alpha \beta} 
P_{\psi}$ is not the identity ${\bf 1}_{\cal H}$ (as claimed in 
\cite{slavedirac}), but the projector $P_{\psi}$ on the $k$-dimensional 
image of $\psi$ in the $k ( N+1 )$-dimensional space ${\cal H}$. 
Thereby physical combinations $\psi_{\alpha}^{\dagger} \psi_{\beta} =
{\bf 1}_k \otimes D_{\alpha \beta}$ 
with ${ ( D_{\alpha \beta} )}_{0 \le n,m \le N} 
 = \delta_{n \alpha} \delta_{m \beta}$
 representing $X_{\alpha \beta}$ are blockdiagonal
in ${\cal H}$ and  the expectation values of all {\it physical}
operators $A_{\cal H} = {\bf 1}_k \otimes A_{{\cal H}_{\rm phys}}$,
which are functions of $\psi_{\alpha}^{\dagger} \psi_{\beta}$ only,
are independent of the choice of the representation and coincide with the
physical result:
\begin{eqnarray} \label{expectation}
{ \langle A ( \psi^{\dagger}_{\alpha} \psi_{\beta}) \rangle }_{\cal H} & = &
\frac{{\rm tr}_{\cal H} (e^{- \beta H}
 A )}{{\rm tr}_{\cal H}  ( e^{- \beta H} ) } =
\frac{{\rm tr}_{{\cal H}_{\rm phys}} (e^{- \beta H}
 A )}{{\rm tr}_{{\cal H}_{\rm phys}}
  ( e^{- \beta H} ) }  \nonumber        \\  
&=& { \langle A ( X_{\alpha \beta} ) \rangle}_{{\cal H}_{\rm phys}}   . 
\end{eqnarray}

With the help of these auxiliary operators the Laplace transform of the 
f-electron Greenfunction 
\begin{equation}
G_{\alpha \beta} ( t ) : = - i 
  \langle  { [ X_{0 \alpha} (t) , X_{\beta 0 } ] }_{+} \rangle 
  \Theta ( t )
\end{equation}
can be written as 
\begin{equation}
\begin{array}{l}
 G_{\alpha \beta} ( z ) =  \\ 
\int_{\cal C} \frac{d z_1}{2 \pi i } 
  \langle   [ ( \frac{1}{z_1 - {\cal L} } b^{\dagger} ) 
    ( \frac{1}{z + z_1 + {\cal L}} f_{\alpha} ) , 
    f_{\beta}^{\dagger} b ]_{+} \rangle  , 
\end{array}
\end{equation}
${\cal L}$ being the Liouville operator and the contour 
${\cal C}$ enclosing all eigenvalues of $H$.
The expansion of $G_{\alpha \beta}$ in orders of $V$ using  the formula 
\begin{equation}
\frac{1}{z - {\cal L}_0  - {\cal L}_V } = \frac{1}{z - {\cal L}_0} 
\sum_{n=0}^{\infty} {  ( {\cal L}_V \frac{1}{z - {\cal L}_0} ) }^{n} 
\end{equation}
can  be translated into diagrams similar to those used 
in the usual resolvent theory. 
Thereby the application of ${\cal L}$ on $\psi_{\alpha}^{(\dagger)}$ 
creates unusual non-vanishing  products of four and more slave-operators like 
$\psi_{\alpha}^{\dagger} \psi_{\beta}^{\dagger} \psi_{\gamma} \psi_{\delta}$.

These unphysical terms cancel out in {\it any} representation in any finite 
order in $V$, reproducing the correct results of a direct $X$-operator 
approach order by order, 
which shows the consistency of the approach.
But when summing up an infinite number of diagrams, these unusual terms can 
remain 
and the  results can  depend on the choice of representation.

The approximation $\frac{1}{z -{\cal L}} = \frac{1}{z - {\cal L}_0 - \Sigma }$
with $ \Sigma = {\cal L}_V \frac{1}{z - {\cal L}_0} {\cal L}_V $ 
shows Kondo-like resonances at the energy 
$ T_0 = D {\rm exp} (- \frac{  \mid \epsilon_f \mid}{ ( N-1 ) N_0 V^2 } )  $ 
- consistent with the physical intuition - 
only for spin degeneracy $N>1$. Here the inclusion of vertex corrections 
seems to be crucial for the reproduction of the correct Kondo scale for 
$N \neq 1, \infty$. 
This and the comparison
 with the usual resolvent perturbation theory, which should reveal the 
physical significance of the used subclass of diagrams, are currently 
being investigated. 


To conclude, we constructed new  slave-particle decompositions of the 
Hubbard-operators, which are {\it exact} on the 
operator level, and developed a formal concept for a  perturbation 
theory  of these auxiliary operators. This will allow to study the
 above theory in (e.g. selfconsistent) approximations, which go beyond the 
results presented above.

One of the authors (C.H.) of this work has been supported by grants 
from  the Studienstiftung des Deutschen Volkes.

\setlength{\evensidemargin}{10mm}
\setlength{\oddsidemargin}{10mm}

\begin{minipage}[t]{14cm}

\end{minipage}

\end{document}